\documentclass[a4paper,aps,pra,twocolumn,showpacs,amsmath,amssymb]{revtex4}
\usepackage{amsmath}
\usepackage{graphicx}
\newcommand{\ud}{\!\!\mathrm{d}}
\newcommand{\parder}[1]{\partial_{#1}}
\newcommand{\p}[2]{p_\mathrm{\,#1}^\mathrm{\,#2}}
\newcommand{\x}[2]{x_\mathrm{\,#1}^\mathrm{\,#2}}
\newcommand{\xa}[1]{X^{\,\mathrm{#1}}}
\newcommand{\pa}[1]{P^{\,\mathrm{#1}}}
\newcommand{\y}[1]{y_\mathrm{\,#1}}
\newcommand{\q}[1]{q_\mathrm{\,#1}}

\begin{document}

\title{Teleportation and spin squeezing utilizing multimode entanglement of light with atoms}

\author{K. Hammerer$^1$, E.S. Polzik$^{2,3}$, J.I. Cirac$^1$}

\affiliation{ $^1$Max-Planck--Institut f\"ur Quantenoptik,
Hans-Kopfermann-Strasse, D-85748 Garching, Germany \\
$^2$ QUANTOP, Danish Research Foundation Center for Quantum Optics, DK 2100 Copenhagen, Denmark\\
$^3$ Niels Bohr Institute, DK 2100 Copenhagen, Denmark}

\begin{abstract}
We present a protocol for the teleportation of the quantum state
of a pulse of light onto the collective spin state of an atomic
ensemble. The entangled state of light and atoms employed as a
resource in this protocol is created by probing the collective
atomic spin, Larmor precessing in an external magnetic field, off
resonantly with a coherent pulse of light. We take here for the
first time full account of the effects of Larmor precession and
show that it gives rise to a qualitatively new type of multimode
entangled state of light and atoms. The protocol is shown to be
robust against the dominating sources of noise and can be
implemented with an atomic ensemble at room temperature
interacting with free space light. We also provide a scheme to
perform the readout of the Larmor precessing spin state enabling
the verification of successful teleportation as well as the
creation of spin squeezing.
\end{abstract}

\pacs{03.67.Mn, 32.80.Qk}

\maketitle


\section{Introduction}

Quantum teleportation - the disembodied transport of quantum
states - has been demonstrated so far in several seminal
experiments dealing with purely photonic \cite{PHOTONS} or atomic
\cite{ATOMS} systems. Here we propose a protocol for the
teleportation of a coherent state carried initially by a pulse of
light onto the collective spin state of $\sim\!10^{11}$ atoms.
This protocol - just as the recently demonstrated direct transfer
of a quantum state of light onto atoms \cite{JSCFP} - is
particularly relevant for long distance entanglement distribution,
a key resource in quantum communication networks \cite{BP}.

Our scheme can be implemented with just coherent light and
room-temperature atoms in a \textit{single} vapor cell placed in a
homogeneous magnetic field. Existing protocols in Quantum
Information (QI) with continuous variables of atomic ensembles and
light \cite{BP} are commonly designed for setups where no external
magnetic field is applied such that the interaction of light with
atoms meets the Quantum non-demolition (QND) criteria \cite{HCWL,
PRG}. In contrast, in all experiments dealing with vapor cells at
room-temperature \cite{JSCFP, JKP} it is, for technical reasons,
absolutely essential to employ magnetic fields. In experiments
\cite{JSCFP, JKP} two cells with counter-rotating atomic spins
were used to comply with both, the need for an external magnetic
field and the one for an interaction of QND character. So far it
was believed to be impossible to use a single cell in a magnetic
field to implement QI protocols, since in this case -  due to the
Larmor precession -  scattered light simultaneously reads out two
non-commuting spin components such that the interaction is not of
QND type.

In this paper we do not only show that it is well possible to make
use of the quantum state of light and atoms created in this setup
but we demonstrate that - for the purpose of teleportation
\cite{V, BK} - it is in fact better to do so. As compared to the
state resulting from the common QND interaction the application of
an external magnetic field enhances the creation of correlations
between atoms and light, generating more and qualitatively new,
multimode type of entanglement. The results of the paper can be
summarized as follows:\\
(\textit{i}) Larmor precession in an external magnetic field
enhances the creation of entanglement when a collective atomic
spin is probed with off-resonant light. The resulting entanglement
involves multiple modes and is stronger as compared to what can be
achieved in a comparable QND interaction.\\
(\textit{ii}) This type of entangled state can be used as a
resource in a teleportation protocol, which is a simple
generalization of the standard protocol \cite{V, BK} based on
Einstein-Podolsky-Rosen (EPR) type of entanglement. For the
experimentally accessible parameter regime the teleportation
fidelity is close to optimal. The protocol is robust against
imperfections and can be implemented with state of the art
technique.\\
(\textit{iii}) Homodyne detection of appropriate scattering modes
of light leaves the atomic state in a spin squeezed state. The
squeezing can be the same as attained from a comparable QND
measurement of the atomic spin \cite{KMB, GSM}. The same scheme
can be used for atomic state read-out of the Larmor precessing
spin, necessary to verify successful teleportation.\\

We would like to note that it was shown recently in \cite{OF} that
the effect of a magnetic field can enhance the capacity of a
quantum memory in the setup of two cells. Teleportation in the
setup of a single cell without magnetic field was addressed in
\cite{MF}.

The paper is organized as follows: The three points above are
presented in sections \ref{interaction}, \ref{teleportation} and
\ref{spinsqu&readout}, in this order. Some of the details in the
calculations of sections \ref{teleportation} and
\ref{spinsqu&readout} are moved to appendices \ref{backaction} and
\ref{feedback}.


\section{Interaction}\label{interaction}
We consider an ensemble of $N_{at}$ Alkali atoms with total ground
state angular momentum $F$, placed in a constant magnetic field
causing a Zeeman splitting of $\hbar\Omega$ and initially prepared
in a fully polarized state along $x$. The collective spin of the
ensemble is then probed by an off resonant pulse which propagates
along $z$ and is linearly polarized along $x$. Thorough
descriptions of this interaction and the final state of light and
atoms after the scattering can be found in \cite{KMPP, KBM, TMW,
DCZP, KMP} and especially in \cite{JSSP, J, SJP, KMSJP} for the
specific system we have in mind. We derive the final state here
with a special focus on the effects of Larmor precession and light
propagation in order to identify the light modes which are
actually populated in the scattering process.

In appendix \ref{appendix} we show that the interaction is
adequately described by a Hamiltonian
\begin{eqnarray}\label{Hamiltonian}
H&=&H_{at}+H_{li}+V,\nonumber\\
H_{at}&=&\frac{\hbar\Omega}{2}(X^2+P^2),\\
V&=&\frac{\hbar\kappa}{\sqrt{T}} Pp(0)\nonumber
\end{eqnarray}
and where $H_{li}$ is the Hamiltonian for the free radiation
field. The canonical conjugate variables $X,P$ describe in the
Holstein-Primakoff approximation \cite{K} transverse components of
the collective angular momentum in $y$ and $z$ direction
respectively. They satisfy $[X,P]=i$ and have zero mean and a
normalized variance \mbox{$\Delta X^2=\Delta P^2=1/2$} for the
initial coherent spin state. In analogy to light field quadratures
we will denote the normalized transverse spin components $X,\,P$
also as spin quadratures. For the light field only its linearly
polarized component along $y$ is relevant and is described in
terms of quadratures of spatially localized modes \cite{SD, MM},
$x(z),p(z)$, which obey \mbox{$[x(z),p(z')]=ic\delta(z-z')$}.
Before the interaction process, this polarization component is in
vacuum such that initially \mbox{$\langle x(z)\rangle=\langle
p(z)\rangle=0$} and \mbox{$\langle x(z)x(z')\rangle=\langle
p(z)p(z')\rangle=c\delta(z-z')/2$}. The dimensionless coupling
constant is given by
\mbox{$\kappa=\sqrt{N_{ph}N_{at}F}a_1\sigma\Gamma/2A\Delta$} where
$N_{ph}$ is the overall number of photons in the pulse, $a_1$ is a
constant characterizing the ground state's vector polarizability,
$\sigma$ is the scattering cross section, $\Gamma$ the decay rate,
$\Delta$ the detuning and $A$ the effective beam cross section.

Changing to a rotating frame with respect to $H_{at}$ by defining
$X_I(t)=\exp(-iH_{at}t)X\exp(iH_{at}t)$ and evaluating the
Heisenberg equations for these operators yields the following
Maxwell-Bloch equations
\begin{subequations}
\begin{eqnarray}
\parder{t}X_I(t)\!\!&=&\!\!\frac{\kappa}{\sqrt{T}}\cos(\Omega t)p(0,t),\label{X}\\
\parder{t}P_I(t)\!\!&=&\!\!\frac{\kappa}{\sqrt{T}}\sin(\Omega t)p(0,t),\label{P}\\
\left(\parder{t}\!+\!c\parder{z}\right)x(z,t)\!\!&=&\!\!\frac{\kappa
c }{\sqrt{T}}\left[\cos(\Omega
t)P_I(t)-\sin(\Omega t)X_I(t)\right]\delta(z),\nonumber\\
\left(\parder{t}\!+\!c\parder{z}\right)p(z,t)\!\!&=&\!\!0,\nonumber
\end{eqnarray}
where $\parder{t(z)}$ denotes the partial derivative with respect
to $t\,(z)$. These equations have a clear interpretation. Light
noise coming from the field in quadrature with the classical probe
piles up in both, the $X$ and $P$ spin quadrature, but it
alternately affects only one or the other, changing with a period
of $1/\Omega$. Conversely atomic noise adds to the in phase field
quadrature only and the signal comes alternately from the $X$ and
$P$ spin quadrature. The out of phase field quadrature is
conserved in the interaction.

To solve this set of coupled equations it is convenient to
introduce a new position variable, $\xi=ct-z$, to eliminate the
$z$ dependence. New light quadratures defined by
$\bar{x}(\xi,t)=x(ct-\xi,t),\,\bar{p}(\xi,t)=p(ct-\xi,t)$ also
have a simple interpretation: $\xi$ labels the slices of the pulse
moving in and out of the ensemble one after the other, starting
with $\xi=0$ and terminating at $\xi=cT$. The Maxwell equations
now read

\begin{align}
&\parder{t}\bar{p}(\xi,t)=0,\label{pbar}\\
&\parder{t}\bar{x}(\xi,t)=\frac{\kappa c
}{\sqrt{T}}\left[\cos(\Omega t)P_I(t)-\sin(\Omega
t)X_I(t)\right]\delta(ct-\xi).\label{xbar}
\end{align}
\end{subequations}

The solutions to equations (\ref{X}, \ref{P}, \ref{pbar}) are
\begin{subequations}
\begin{align}
X_I(t)&=X_I(0)+\frac{\kappa}{\sqrt{T}}\int_0^t\ud\tau\cos(\Omega\tau)\bar{p}(c\tau,0),\label{Xsol}\\
P_I(t)&=P_I(0)+\frac{\kappa}{\sqrt{T}}\int_0^t\ud\tau\sin(\Omega\tau)\bar{p}(c\tau,0),\label{Psol}\\
\bar{p}(\xi,t)&=\bar{p}(\xi,0)\\
\intertext{and the formal solution to \eqref{xbar} is}
\bar{x}(\xi,t)&=\bar{x}(\xi,0)+\label{xlight}\\
&+\frac{\kappa}{\sqrt{T}}\left[\cos(\Omega\xi/c)P_I(\xi/c)-\sin(\Omega\xi/c)X_I(\xi/c)\right].\nonumber
\end{align}
\end{subequations}

As mentioned before, both atomic spin quadratures are affected by
light but, as is evident from the solutions for $X(t),\,P(t)$,
they receive contributions from different and, in fact, orthogonal
projections of the out-of-phase field. As we will show in the
following, the corresponding projections of the in-phase field
carry in turn the signal of atomic quadratures after the
interaction. It is therefore convenient to explicitly introduce
operators for these modes \cite{SJP}. We define a cosine component
before the interaction
\begin{subequations}\label{cosine}
\begin{eqnarray}
\p{c}{in}&=&\sqrt{\frac{2}{T}}\int_0^T\ud\tau\cos(\Omega\tau)\bar{p}(c\tau,0),\\
\x{c}{in}&=&\sqrt{\frac{2}{T}}\int_0^T\ud\tau\cos(\Omega\tau)\bar{x}(c\tau,0)
\end{eqnarray}
\end{subequations}
and a sine component $\p{s}{in},\,\x{s}{in}$ with
$\cos(\Omega\tau)$ replaced by $\sin(\Omega\tau)$. In frequency
space these modes consist of spectral components at sidebands
$\omega_c\pm\Omega$ and are closely related to the sideband
modulation modes introduced in \cite{CS} for the description of
two photon processes. It is easily checked that these modes are
asymptotically canonical,
\mbox{$[\x{c}{in},\p{c}{in}]=[\x{s}{in},\p{s}{in}]=i[1+\mathcal{O}(n_0^{-1})]\simeq
i$}, and independent,
$[\x{c}{in},\p{s}{in}]=\mathcal{O}(n_0^{-1})\simeq 0$, if we
assume $n_0\gg 1$ for \mbox{$n_0=\Omega T$}, the pulse length
measured in periods of Larmor precession.

In terms of these modes the atomic state after the interaction
\mbox{$\xa{out}=X_I(T),\,\pa{out}=P_I(T)$} is given by
\begin{subequations}\label{finalstate}
\begin{equation}
\xa{out}=\xa{in}+\frac{\kappa}{\sqrt{2}}\p{c}{in},\quad\pa{out}=\pa{in}+\frac{\kappa}{\sqrt{2}}\p{s}{in}.\label{finalstateatoms}\\
\end{equation}
The final state of cosine (sine) modes is described by
$\x{c(s)}{out},\,\p{c(s)}{out}$, defined by equations
(\ref{cosine}) with $\bar{x}(c\tau,0),\,\bar{p}(c\tau,0)$ replaced
by $\bar{x}(c\tau,T),\,\bar{p}(c\tau,T)$ respectively. Since the
out-of-phase field is conserved we have trivially
\begin{equation}
\p{c}{out}=\p{c}{in},\quad\p{s}{out}=\p{s}{in}.
\end{equation}
Deriving the corresponding expressions for the cosine and sine
components of the field in phase, $\x{c}{out},\,\x{s}{out}$,
raises some difficulties connected to the back action of light
onto itself. This effect can be understood by noting that a slice
$\xi$ of the pulse receives a signal of atoms at a time $\xi/c$
[see equation (\ref{xlight})] which, regarding equations
(\ref{Xsol}, \ref{Psol}), in turn carry already the integrated
signal of all slices up to $\xi$. Thus, mediated by the atoms,
light acts back on itself. The technicalities in the treatment of
this effect are given in appendix \ref{backaction} where we
identify relevant "back action modes",
$\x{c,1}{},\,\p{c,1}{},\,\x{s,1}{},\,\p{s,1}{}$, in terms of which
one can express
\begin{eqnarray}
\x{c}{out}=\x{c}{in}\!+\!\frac{\kappa}{\sqrt{2}}\pa{in}\!+\!\left(\frac{\kappa}{2}\right)^2\p{s}{in}\!
+\!\frac{1}{\sqrt{3}}\left(\frac{\kappa}{2}\right)^2\p{s,1}{in},\quad&&\label{xcosout}\\
\x{s}{out}=\x{s}{in}\!-\!\frac{\kappa}{\sqrt{2}}\xa{in}\!-\!\left(\frac{\kappa}{2}\right)^2\p{c}{in}\!
-\!\frac{1}{\sqrt{3}}\left(\frac{\kappa}{2}\right)^2\p{c,1}{in}.\quad&&\label{xsinout}
\end{eqnarray}
\end{subequations}
The last two terms in both lines represent the effect of back
action, part of which involves the already defined cosine and sine
components of the field in quadrature. The remaining part is
subsumed in the back action modes which are again canonical and
independent from all other modes.

\begin{figure}
\includegraphics[width=6.5cm]{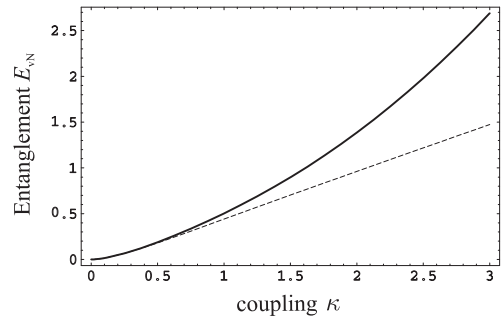}
\caption{Von Neumann Entropy of the reduced state of atoms versus
coupling strength $kappa$ for the state of equation
\ref{finalstate} (full line) and for the state generated without
magnetic field in a pure QND interaction (dashed line) with the
same coupling strength. Application of a magnetic field
significantly enhances the amount of light-atom
entanglement.}\label{EvNvsKappa}
\end{figure}

Equations (\ref{finalstate}) describe the final state of atoms and
the relevant part of scattered light after the pulse has passed
the atomic ensemble and are the central result of this section.
Treating the last terms in equations (\ref{xcosout},\ref{xsinout})
as noise terms, it is readily checked by means of the separability
criteria in \cite{GKLC} that this state is fully inseparable,
\textit{i.e.} it is inseparable with respect to all splittings
between the three modes. For the following teleportation protocol
the relevant entanglement is the one between atoms and the two
light modes. Figure \ref{EvNvsKappa} shows the von Neumann entropy
$E_{\mathrm{vN}}$ of the reduced state of atoms in its dependence
on the coupling strength $\kappa$ and in comparison with the
entanglement created without magnetic field in a pure QND
interaction of atoms and light. The amount of entanglement is
significantly enhanced.


\section{Teleportation of light onto atoms}\label{teleportation}

In this section we will show how the multimode entanglement
between light and atoms generated in the scattering process can be
employed in a teleportation protocol which is a simple
generalization of the standard protocol for continuous variable
teleportation using EPR-type entangled states \cite{V, BK}. We
first present the protocol and evaluate its fidelity and then
analyze its performance under realistic experimental conditions.

\subsection{Basic protocol}\label{protocol}

\begin{figure}
\includegraphics[width=8cm]{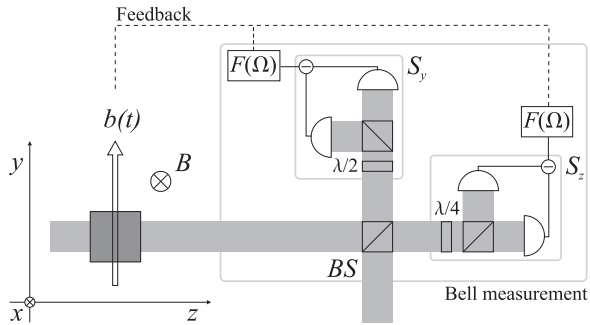}
\caption{Scheme for teleportation of light onto atoms: As
described in section \ref{interaction}, a classical pulse
(linearly polarized along $x$) propagating along the positive $z$
direction is scattered off an atomic ensemble contained in a glass
cell and placed in a constant magnetic field $B$ along $x$.
Classical pulse and scattered light (linearly polarized along $y$)
are overlapped with a with a coherent pulse (linearly polarized
along $z$) at beam splitter $B\!S$. By means of standard
polarization measurements Stokes vector components $S_y$ and $S_z$
are measured at one and the other port respectively, realizing the
Bell measurement. The Fourier components at Larmor frequency
$\Omega$ of the corresponding photocurrents determine the amount
of conditional displacement of the atomic spin which can be
achieved by applying a properly timed transverse magnetic field
$b(t)$. See section \ref{protocol} for details.}\label{scheme}
\end{figure}

Figure \ref{scheme} depicts the basic scheme which, as usually,
consists of a Bell measurement and a feedback operation.

\paragraph*{Input} The coherent state to be teleported is encoded in a pulse which
is linearly polarized orthogonal to the classical driving pulse
and whose carrier frequency lies at the upper sideband, i.e. at
$\omega_c+\Omega$. The pulse envelope has to match the one of the
classical pulse. As is shown in appendix \ref{backaction},
canonical operators $y,q$ with $[y,q]=i$ describing this mode can
conveniently be expressed in terms of cosine and sine modulation
modes, analogous to equations (\ref{cosine}), defined with respect
to the carrier frequency. One finds
\begin{equation}\label{inputstate}
y=\frac{1}{\sqrt{2}}\left(\y{s}+\q{c}\right),\quad
q=-\frac{1}{\sqrt{2}}\left(\y{c}-\q{s}\right).
\end{equation}
A coherent input amounts to having initially \mbox{$\Delta
y^2=\Delta q^2=1/2$} and an amplitude $\langle y\rangle,\,\langle
q\rangle$ with mean photon number $n_{ph}=(\langle
y\rangle^2+\langle q\rangle^2)/2$.

\paragraph*{Bell measurement} This input is combined at a beam splitter
with the classical pulse and the scattered light. At the ports of
the beam splitter Stokes vector components $S_y$ and $S_z$ are
measured by means of standard polarization measurements. Given the
classical pulse in $x$ polarization this amounts to a homodyne
detection of in- and out-of-phase fields of the orthogonal
polarization component. The resulting photocurrents are
numerically demodulated to extract the relevant sine and cosine
components at the Larmor frequency \cite{J}. Thus one effectively
measures the commuting observables
\begin{align}
\tilde{x}_{\mathrm{c}}&=\frac{1}{\sqrt{2}}\left(\x{c}{out}+\y{c}\right),
&\tilde{x}_{\mathrm{s}}&=\frac{1}{\sqrt{2}}\left(\x{s}{out}+\y{s}\right),\label{measurement}\raisetag{-0pt}\\
\tilde{q}_{\mathrm{c}}&=\frac{1}{\sqrt{2}}\left(\p{c}{out}-\q{c}\right),
&\tilde{q}_{\mathrm{s}}&=\frac{1}{\sqrt{2}}\left(\p{s}{out}-\q{s}\right).\nonumber
\end{align}
Let the respective measurement results be given by
$\tilde{X}_{\mathrm{c}},\,\tilde{X}_{\mathrm{s}},\,\tilde{Q}_{\mathrm{c}}$
and $\tilde{Q}_{\mathrm{s}}$.

\paragraph*{Feedback} Conditioned on these results the atomic state is then
displaced by an amount
\mbox{$\tilde{X}_{\mathrm{s}}-\tilde{Q}_{\mathrm{c}}$} in $X$ and
\mbox{$-\tilde{X}_{\mathrm{c}}-\tilde{Q}_{\mathrm{s}}$} in $P$.
This can be achieved by means of two fast radio-frequency magnetic
pulses separated by a quarter of a Larmor period. In the ensemble
average the final state of atoms is simply given by
\begin{equation}\label{finalstateatoms1}
\xa{fin}=\xa{out}+\tilde{x}_{\mathrm{s}}-\tilde{q}_{\mathrm{c}},\quad
\pa{fin}=\pa{out}-\tilde{x}_{\mathrm{c}}-\tilde{q}_{\mathrm{s}}.
\end{equation}
This description of feedback is justified rigorously in appendix
\ref{feedback}. Relating these expressions to input operators, we
find by means of equations \eqref{finalstate}, \eqref{inputstate}
and \eqref{measurement}
\begin{subequations}\label{finalstateatoms2}
\begin{eqnarray}
\xa{fin}&=&\left(1-\frac{\kappa}{2}\right)\xa{in}-\frac{1}{\sqrt{2}}\left(1-\frac{\kappa}{2}\right)^2\p{c}{in}\nonumber\\
&&+\frac{1}{\sqrt{2}}\x{s}{in}-\frac{1}{\sqrt{6}}\left(\frac{\kappa}{2}\right)^2\p{c,1}{in}+y,\\
\pa{fin}&=&\left(1-\frac{\kappa}{2}\right)\pa{in}-\frac{1}{\sqrt{2}}\left(1-\frac{\kappa}{2}\right)^2\p{s}{in}\nonumber\\
&&-\frac{1}{\sqrt{2}}\x{c}{in}-\frac{1}{\sqrt{6}}\left(\frac{\kappa}{2}\right)^2\p{s,1}{in}+q.
\end{eqnarray}
\end{subequations}
This is the main result of this section.

\begin{figure}
\includegraphics[width=6.5cm]{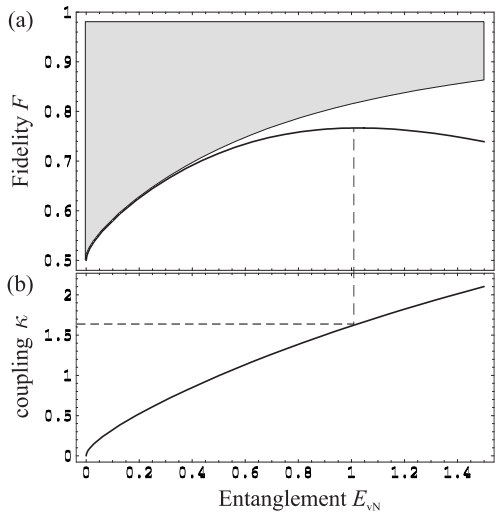}
\caption{(a) Theoretical limit on the achievable fidelity $F$
versus entanglement between atoms and light measured by the von
Neumann entropy $E_{\mathrm{vN}}$ of the reduced state of atoms.
The grey area is unphysical. For moderate amounts of entanglement
our protocol is close to optimal. (b) Coupling strength $\kappa$
versus entanglement. The dashed lines indicate the maximal
fidelity of $F=.77$ which is achieved for $\kappa=1.64$.}
\label{fidelity}
\end{figure}

\paragraph*{Teleportation fidelity} Taking the mean of equations
\eqref{finalstateatoms2} with respect to the initial state all
contributions due to input operators and back action modes vanish
such that $\langle\xa{fin}\rangle=\langle y\rangle$ and
$\langle\pa{fin}\rangle=\langle q\rangle$. Thus, the amplitude of
the coherent input light pulse is mapped on atomic spin
quadratures as desired. In order to prove faithful teleportation
also the variances have to be conserved. It is evident from
\eqref{finalstateatoms2} that the final atomic spin variances will
be increased as compared to the coherent input. These additional
terms describe unwanted excess noise and have to be minimized by a
proper choice of the coupling $\kappa$. As a figure of merit for
the teleportation protocol we use the fidelity, i.e. squared
overlap, of input and final state. Given that the means are
transmitted correctly the fidelity is found to be
$F=2\left[(1+2(\Delta\xa{fin})^2)(1+2(\Delta\pa{fin})^2)\right]^{-1/2}$.
The variances of the final spin quadratures are readily calculated
taking into account that all modes involved are independent and
have initially a normalized variance of 1/2. In this way a
theoretical limit on the achievable fidelity can be derived
depending solely on the coupling strength $\kappa$. In figure
\ref{fidelity} we take advantage of the fact that the amount of
entanglement between light and atoms is a monotonously increasing
function of $\kappa$ such that we can plot the fidelity versus the
entanglement. This has the advantage that we can compare the
performance of our teleportation protocol with the canonical one
\cite{V, BK} which uses a two-mode squeezed state of the same
entanglement as a resource and therefore maximizes the
teleportation fidelity for the given amount of entanglement. No
physical state can achieve a higher fidelity with the same
entanglement. This follows from the results of \cite{GWKWC} where
it was shown that two-mode squeezed states minimize the EPR
variance (and therefore maximize the teleportation fidelity) for
given entanglement. The theoretical fidelity achievable in our
protocol is maximized for $\kappa\simeq 1.64$ corresponding to
$F\simeq.77$. But also for experimentally more feasible values of
$\kappa\simeq 1$ can the fidelity well exceed the classical limit
\cite{BFK, HWPC} of 1/2 and, moreover, comparison with the values
achievable with a two-mode squeezed state shows that our protocol
is close to optimal.

\subsection{Noise effects and Gaussian distributed
input}\label{noise}

Under realistic conditions the teleportation fidelity will be
degraded by noise effects like decoherence of the atomic spin
state, light absorption and reflection losses and also because the
coupling constant $\kappa$ is experimentally limited to values
$\kappa\simeq 1$. On the other hand the classical fidelity bound
to be beaten will be somewhat higher than 1/2 since the coherent
input states will necessarily be drawn according to a distribution
with a finite width in the mean photon number $\bar{n}$. In this
section we analyze the efficiency of the teleportation protocol
under these conditions and show that it is still possible to
surpass any classical strategy for the transmission and storage of
coherent states of light \cite{BFK, HWPC}.

During the interaction atomic polarization decays due to
spontaneous emission and collisional relaxation. Including a
transverse decay the final state of atoms is given by
\begin{subequations}\label{noisyatomstate}
\begin{eqnarray}
\xa{out}&=&\sqrt{1-\beta}(\xa{in}+\frac{\kappa}{\sqrt{2}}\p{c}{in})+\sqrt{\beta}f_X,\\
\pa{out}&=&\sqrt{1-\beta}(\pa{in}+\frac{\kappa}{\sqrt{2}}\p{s}{in})+\sqrt{\beta}f_P.
\end{eqnarray}
\end{subequations}
as follows from the discussion in appendix \ref{appendix}. $\beta$
is the atomic decay parameter and $f_X,\,f_P$ are Langevin noise
operators with zero mean. Their variance is experimentally found
to be close to the value corresponding to a coherent state such
that \mbox{$\langle f_X^{\,2}\rangle=\langle
f_P^{\,2}\rangle=1/2$}.

Light absorption and reflection losses can be taken into account
in the same way as finite detection efficiency. For example the
statistics of measurement outcome $\tilde{X}_{\mathrm{s}}$ will
not stem from the signal mode $\tilde{x}_{\mathrm{s}}$ alone but
rather from the noisy mode
\mbox{$\sqrt{1-\epsilon}\,\tilde{x}_{\mathrm{s}}+\sqrt{\epsilon}f_{x,\mathrm{s}}$}
where $\epsilon$ is the photon loss parameter and
$f_{x,\mathrm{s}}$ is a Langevin noise operator of zero mean and
variance $\langle f_{x,\mathrm{s}}^{\,2}\rangle=1/2$. Analogous
expressions have to be used for the measurements of
$\tilde{x}_{\mathrm{c}},\,\tilde{q}_{\mathrm{s}}$ and
$\tilde{q}_{\mathrm{c}}$ which will be adulterated by Langevin
terms $f_{x,\mathrm{c}},\,f_{q,\mathrm{s}}$ and $f_{q,\mathrm{c}}$
respectively. In principle each of the measurement outcomes can be
fed back with an independently chosen gain but for symmetry
reasons it is enough to distinguish gain coefficients $g_x,\,g_q$
for the measurement outcomes of sine and cosine components of $x$
and $q$ respectively. Including photon loss, finite gain and
atomic decay, as given in \eqref{noisyatomstate}, equations
\eqref{finalstateatoms1}, describing the final state of atoms
after the feed back operation, generalize to
\begin{subequations}\label{noisyfinalstate}
\begin{eqnarray}
\xa{fin}&=&\sqrt{1-\beta}\xa{out}+\sqrt{\beta}f_X\nonumber\\
&&+g_x\left(\sqrt{1-\epsilon}\,\tilde{x}_{\mathrm{s}}+\sqrt{\epsilon}f_{x,\mathrm{s}}\right)\\
&&-g_q\left(\sqrt{1-\epsilon}\,\tilde{q}_{\mathrm{c}}+\sqrt{\epsilon}f_{q,\mathrm{c}}\right),\nonumber\\
\pa{fin}&=&\sqrt{1-\beta}\pa{out}+\sqrt{\beta}f_P\nonumber\\
&&-g_x\left(\sqrt{1-\epsilon}\,\tilde{x}_{\mathrm{c}}+\sqrt{\epsilon}f_{x,\mathrm{c}}\right)\\
&&-g_q\left(\sqrt{1-\epsilon}\,\tilde{q}_{\mathrm{s}}+\sqrt{\epsilon}f_{q,\mathrm{s}}\right).\nonumber
\end{eqnarray}
\end{subequations}

\begin{figure}
\includegraphics[width=6.9cm]{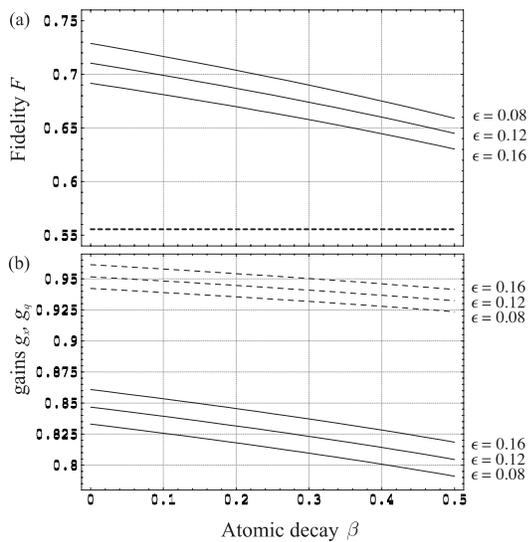}
\caption{(a) Average fidelity achievable in the presence of atomic
decay $\beta$, reflection and light absorption losses
\mbox{$\epsilon=8\%,\,12\%,\,16\%$}, coupling $\kappa=0.96$ and
Gaussian distributed input states with mean photon number
$\bar{n}=4$. The fidelity benchmark is in this case 5/9 (dashed
line). (b) Respective optimal values for gains $g_x$ (solid lines)
and $g_q$ (dashed lines).}\label{noisyfidelity}
\end{figure}

For non unit gains a given coherent amplitude $(\langle
y\rangle,\langle q\rangle)$ will not be perfectly teleported onto
atoms and the corresponding fidelity will be degraded by this
mismatch according to
\begin{eqnarray*}
F(\langle
y\rangle,\langle q\rangle)&=&\frac{2}{\sqrt{[1+2(\Delta\xa{fin})^2][1+2(\Delta\pa{fin})^2]}}\\
&&\cdot\exp\left[-\frac{(\langle
y\rangle-\langle\xa{fin}\rangle)^2}{1+2(\Delta\xa{fin})^2}-\frac{(\langle
q\rangle-\langle\pa{fin}\rangle)^2}{1+2(\Delta\pa{fin})^2}\right].
\end{eqnarray*}
If the input amplitudes are drawn according to a Gaussian
distribution \mbox{$p(\langle y\rangle,\langle
q\rangle)=\exp[-(\langle y\rangle^2+\langle
q\rangle^2)/2\bar{n}]/2\pi\bar{n}$} with mean photon number
$\bar{n}$ the average fidelity [with respect to $(\langle
y\rangle,\langle q\rangle)$] is readily calculated. The exact
expression in terms of initial operators can then be derived by
means of equations \eqref{finalstate}, \eqref{inputstate},
\eqref{measurement} and \eqref{noisyfinalstate} but is not
particularly enlightening. In \mbox{figure \ref{noisyfidelity}} we
plot the average fidelity, optimized with respect to gains
$g_x,\,g_q$, in its dependence on the atomic decay $\beta$ for
various values of photon loss $\epsilon$. We assume a realistic
value $\kappa=0.96$ for the coupling constant and a mean number of
photons $\bar{n}=4$ for the distribution of the coherent input.
For feasible values of $\beta,\,\epsilon\precsim 0.2$ the average
fidelity is still well above the classical bound on the fidelity
\cite{BFK, HWPC}. This proves that the proposed protocol is robust
against the dominating noise effects in this system.

The experimental feasibility of the proposal is illustrated with
the following example. Consider a sample of $N_{at}=10^{12}$
Cesium atoms in a glass cell placed in a constant magnetic field
along the $x$-direction causing a Zeeman splitting of
$\Omega=350\mathrm{\,kHz}$ in the $F=4$ ground state multiplet.
The atoms are pumped into $m_F=4$ and probed on the
\mbox{$\mathrm{D}_2\,(F=4\rightarrow F'=3,4,5)$} transition. The
classical pulse contains an overall number of
$N_{ph}=2.5\,10^{13}$ photons, is detuned to the blue by
$\Delta=1\,\mathrm{GHz}$, has a duration $T=1\,\mathrm{ms}$ and
can have an effective cross section of $A\simeq 6\mathrm{cm}^2$
due to thermal motion of atoms. Under these conditions the tensor
polarizability can be neglected $(\Delta/\omega_\mathrm{hfs}\simeq
10^{-1})$. Also $n_0=\Omega T= 350$ justifies the use of
independent scattering modes. The coupling $\kappa\simeq 1$ and
the depumping of ground state population $\eta\simeq 10^{-1}$ as
desired.

\section{Spin squeezing and state read-out}\label{spinsqu&readout}

In this section we present a scheme for reading out either of the
atomic spin components $X,\,P$ by means of a probe pulse
interacting with the atoms in the one way as described in section
\ref{interaction}. The proposed scheme allows one, on the one
hand, to verify successful receipt of the coherent input
subsequent to the teleportation protocol of section
\ref{teleportation} and, on the other hand, enables to generate
spin squeezing if it is performed on a coherent spin state.

It is well known \cite{KBM, GLP} and was demonstrated
experimentally \cite{KMB, GSM} that the pure interaction $V$, as
given in equation \eqref{Hamiltonian}, can be used to perform a
QND measurement of either of the transverse spin components. At
first sight this seems not to be an option in the scenario under
consideration since the local term $H_{at}$, accounting for Larmor
precession, commutes with neither of the spin quadratures such
that the total Hamiltonian does not satisfy the QND criteria
\cite{HCWL, PRG}. As we have shown in section \ref{interaction}
Larmor precession has two effects: Scattered light is correlated
with both transverse components and suffers from back action
mediated by the atoms. Thus, in order to read out a single spin
component one has to overcome both disturbing effects.

Our claim is that this can be achieved by a simultaneous
measurement of \mbox{$\x{c}{out},\,\p{s}{out},\,\p{s,1}{out}$} or
\mbox{$\x{s}{out},\,\p{c}{out},\,\p{c,1}{out}$} if, respectively,
$X$ or $P$ is to be measured. In the following we consider in
particular the former case but everything will hold with
appropriate replacements also for a measurement of $P$.

\begin{figure}
\includegraphics[width=8cm]{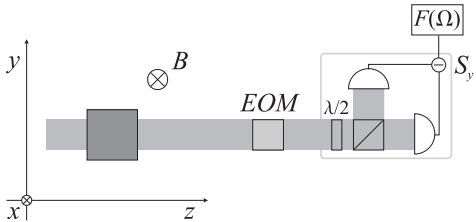}
\caption{Scheme for spin measurement: After the scattering a
$\pi/2$ rotation is performed on the scattered light modulated at
the Larmor frequency such as to affect only the sine (cosine)
component. Standard polarization measurement of $S_y$ and
appropriate postprocessing allows to read out the mean of
$X\,(P)$, leaving the atoms eventually in a spin squeezed state.
}\label{readout}
\end{figure}

As shown in figure \ref{readout} the set of observables
\mbox{$\x{c}{out},\,\p{s}{out},\,\p{s,1}{out}$} can be measured
simultaneously by a measurement of Stokes component $S_y$ after a
$\pi/2$ rotation is performed selectively on the sine component of
the scattered light. The cosine component of the corresponding
photocurrent will give an estimate of $\x{c}{out}$ and the sine
component of $\p{s}{out}$. Multiplying the photocurrent's sine
component by the linear function defining the back action mode,
equation \eqref{back}, will give in addition an estimate of
$\p{s,1}{out}$. Note that the field out of phase is conserved in
the interaction such that
\begin{equation}\label{pback}
\mbox{$\p{s,1}{out}=\p{s,1}{in}$},\quad\mbox{$\p{c,1}{out}=\p{c,1}{in}$},
\end{equation}
i.e. the results will have shot noise limited variance. It is then
evident from equation (\ref{xcosout}) that the respective
photocurrents together with an a priori knowledge of $\kappa$ are
sufficient to estimate the mean $\langle X\rangle$.

The conditional variances after the indicated measurements are
\begin{subequations}\label{condvar}
\begin{eqnarray}
\Delta X^2|_{\{\x{c}{out},\,\p{s}{out},\,\p{s,1}{out}\}}=(\Delta
X^{\mathrm{in}})^2\frac{2}{2+\kappa^2},\\
\Delta P^2|_{\{\x{c}{out},\,\p{s}{out},\,\p{s,1}{out}\}}=(\Delta
P^{\mathrm{in}})^2\frac{2+\kappa^2}{2},
\end{eqnarray}
\end{subequations}
corresponding to a pure state. Obviously the variance in $X$ is
squeezed by a factor \mbox{$(1+\kappa^2/2)^{-1}$}. Note that the
squeezing achieved in a QND measurement without magnetic field but
otherwise identical parameters is given by
\mbox{$(1+\kappa^2)^{-1}$}. From this we conclude that the quality
of the estimate for $\langle X\rangle$, as measured f.e. by
input-output coefficients known from the theory of QND
measurements \cite{HCWL, PRG}, can be the same as in the case
without Larmor precession albeit only for a higher coupling
$\kappa$.

Equations \eqref{condvar} are conveniently derived by means of the
formalism of correlation matrices \cite{GC}. For the operator
valued vector
$\vec{R}=(X,P,\linebreak\x{c}{},\p{c}{},\x{s}{},\p{s}{},\x{c,1}{},\p{c,1}{},\x{s,1}{},\p{s,1}{})$
equations \eqref{finalstate}, \eqref{pback} and \eqref{xback}
define via $\vec{R}^{\mathrm{out}}=S(\kappa)\vec{R}^{\mathrm{in}}$
a symplectic linear transformation $S(\kappa)$. The contributions
of $\p{c,2}{in}$ and $\p{s,2}{in}$ to $\x{s,1}{out}$ and
$\x{c,1}{out}$ as given in \eqref{xback} are treated as noise and
do not contribute to the symplectic transformation $S$ but enter
the input-output relation for the correlation matrix as an
additional noise term as follows. The correlation matrix is as
usually defined by
$\gamma_{i,j}=\mathrm{tr}\{\rho(R_iR_j+R_jR_i)\}$. The initial
state is then an $10\times 10$ identity matrix and the final state
is
$\gamma^\mathrm{out}=S(\kappa)S(\kappa)^T+\gamma_\mathrm{noise}$
where the diagonal matrix
\mbox{$\gamma_\mathrm{noise}=\mathrm{diag}[0,0,0,0,0,0,1,0,1,0](\kappa/2)^4/15$}
accounts for noise contributions due correlations to second order
back action modes c.f. equations \eqref{xback}. In order to
evaluate the atomic variances after a measurement of
\mbox{$\x{c}{out},\,\p{s}{out},\,\p{s,1}{out}$} the correlation
matrix $\gamma_\mathrm{out}$ is split up into blocks,
$$
\gamma_\mathrm{out}=\left(\begin{array}{cc}A & C \\C^{T} &
B\end{array}\right)
$$
where $A$ is the $2\times2$ subblock describing atomic variances.
Now, the state $A'$ after the measurement can be found by
evaluating \cite{GC}
$$
A'=A-\lim_{x,n\rightarrow\infty}C\frac{1}{\Gamma+B}C^T
$$
where $\Gamma=\mathrm{diag}[1/x,x,x,1/x,n,n,x,1/x]$ corresponds to
the measured state. Note that the limit $n\rightarrow\infty$, i.e.
the projection of the unobserved mode $\x{c,2}{},\,\p{c,2}{}$ onto
the identity, does not need to be taken explicitly since,
remarkably, the atomic state after the measurement decouples form
this mode. The conditional variances in equation \eqref{condvar}
are then just (half the) diagonal entries of $A'$.

\section{Conclusions}

In conclusion we have presented a simple and realistic protocol
for teleportation of a coherent state, carried by a propagating
pulse of light, onto the collective spin of an atomic ensemble, a
suitable stationary carrier of quantum information of continuous
variables. The scheme can be implemented with state of the art
technique and allows to surpass any classical strategy for the
transmission and storage of coherent states under realistic
experimental conditions. The basic resource in the protocol is a
multimode entangled state as it results form the interaction of
light with atoms in the presence of an external magnetic field. We
showed that Larmor precession enhances the creation of
entanglement quantitatively and qualitatively. Though the
interaction is not of QND type it is still possible to perform a
state readout on the atomic spin as well as to create significant
spin squeezing. We expect that a proper tailoring of the Larmor
rotation with time dependent magnetic fields would open up
interesting possibilities to further enhance the creation of
entanglement and to deliberately shape scattering modes.

We acknowledge funding from the EU under project
FP6-511004-COVAQIAL and support from \mbox{Kompetenznetzwerk}
Quanteninformationsverarbeitung der Bayerischen Staatsregierung.


\appendix

\section{Hamiltonian}\label{appendix}

In this appendix we present a short derivation of the basic
Hamiltonian \eqref{Hamiltonian} in order to introduce the notation
used throughout this paper. More detailed descriptions can be
found in \cite{KMPP, KBM, TMW, DCZP, KMP, JSSP, J, SJP}. The
Hamiltonian of the system is given by \mbox{$H=H_{at}+H_{li}+V$}
where the atomic part \mbox{$H_{at}=\hbar\Omega\sum_i F_x^{(i)}$}
accounts for the external magnetic field along $x$ causing a
ground state Zeemann splitting of $\hbar\Omega$, $H_{li}$ is the
free space Hamiltonian for light and the interaction term $V$ is
the level shift operator \cite{H, KMSJP}
\begin{equation}\label{levelshift}
V=\int\ud^3 r
\vec{E}^{(-)}(\vec{r}\,)\tensor{\alpha}(\vec{r}\,)\vec{E}^{(+)}(\vec{r}\,),
\end{equation}
which appropriately describes the interaction of off-resonant
light with atoms. We assume here implicitly that the electric
field contains only frequency components within a bandwidth $b$
around the carrier frequency $\omega_c$ of the off-resonant
coherent probe pulse satisfying $\Omega\ll b\ll\Delta_{F'}$ where
the detuning $\Delta_{F'}=\omega_c-\omega_{F,F'}$.

The atomic polarizability density tensor introduced in equation
(\ref{levelshift}) is
\begin{equation}\label{polarizability}
\tensor{\alpha}(\vec{r}\,)=\sum_i\tensor{\alpha}^{(i)}\delta(\vec{r}-\vec{r}^{\,(i)})
\end{equation}
where $\vec{r}^{\,(i)}$ is the position of atom $i$. The single
atom ground state polarizability ${\tensor\alpha}^{(i)}$ consists
in general of a scalar, vector and tensor part,
$$
\tensor{\alpha}=d^2\left(\alpha_0\mathbf{1}+\alpha_1\vec{F}\!\times+\alpha_2\tensor{T}\right),
$$
where $d$ is the relevant reduced dipole matrix element of the
probed transition, $\mathbf{1}$ is the $3\times3$ identity matrix
and $\vec{F}\!\times$ has to be understood to give the vector
cross product of $\vec{F}$ with the vector to the right. Each of
the coefficients $\alpha_j$ is a sum of contributions from
transitions to all excited states manifolds $F'$. If the detuning
is much larger than the typical excited states' hyperfine
splitting, $\Delta\gg\omega_\mathrm{hfs}$, one finds that
$\alpha_2\rightarrow 0$ such that the second rank polarizability
can be neglected. In this case
\begin{equation}\label{singlepolarizability}
\tensor{\alpha}=\frac{d^2}{\hbar(\Delta-i\Gamma/2)}\left(a_0\mathbf{1}+i
a_1\vec{F}\!\times\right)
\end{equation}
with real dimensionless coefficients $a_j$ of order unity and
$\Gamma$ the excited states' decay rate. The non-hermitian part of
the resulting Hamilton operator describes the effect of light
absorption and loss of ground state population due to depumping in
the course of interaction. In the following we will focus on the
coherent interaction and, for the time being, take into account
only the hermitian component. The effects of light absorption and
atomic depumping are treated below.

\paragraph*{Coherent interaction} Since scattering of light occurs predominantly in the forward
direction \cite{KMSJP} it is legitimate to adopt a one dimensional
model such that the (negative frequency component of the) electric
field propagating along $z$ is given by
\begin{eqnarray*}
\vec{E}^{(-)}(z,t)&=&E^{(-)}(z)\vec{e}_y+{\cal{E}}^{(-)}(z,t)\vec{e}_x\\
E^{(-)}(z)&=&\rho(\omega_c)\int_b\ud \omega
a^\dagger(\omega)e^{-ikz}\\
{\cal{E}}^{(-)}(z,t)&=&\rho(\omega_c)\sqrt{4\pi
N_{ph}/T}e^{-i(k_cz-\omega_ct)}
\end{eqnarray*}
where $\rho(\omega)=\sqrt{\hbar\omega_c/4\pi\epsilon_0Ac}$ and $A$
denotes the pulse's cross sectional area, $N_{ph}$ the overall
number of photons in the pulse and $T$ its duration. We restrict
the field in $x$ polarization to the classical probe pulse, since
only the coupling of atoms to the $y$ polarization is enhanced by
the coherent probe. Furthermore we implicitly assume for the
classical pulse a slowly varying envelope such that it arrives at
$z=0$ at $t=0$ and is then constant for a time $T$. Combining this
expression for the field with expressions (\ref{polarizability})
and(\ref{singlepolarizability}) for the atomic polarizability in
equation (\ref{levelshift}) yields
\begin{equation*}
V=-\frac{i\hbar\kappa}{\sqrt{4\pi JT}}\int_b\ud\omega\int\ud z
j(z)\left(a(\omega)e^{-i[(k_c-k)z-\omega_ct]}-h.c.\right)
\end{equation*}
where we defined a dimensionless coupling constant
\mbox{$\kappa=\sqrt{N_{ph}J}\omega_ca_1d^2/\hbar\epsilon_0cA\Delta
$} and an atomic spin density
$j_z(z)=\sum_iF^{(i)}_z\delta(z-z^{(i)})$. In this expression we
skipped terms proportional to $a_0$ which will give rise only to a
global phase shift and included for convenience the square root
factor with $J=N_{at}F$ where $N_{at}$ is the number of atoms.
Note that the coupling can be expressed as
\mbox{$\kappa=\sqrt{N_{ph}J}a_1\sigma\Gamma/2A\Delta$} with
$\sigma$ the scattering cross section on resonance.

We now define field quadratures for spatially localized modes
\cite{SD, MM} as
\begin{subequations}\label{localizedmodes}
\begin{eqnarray}
x(z)&=&\frac{1}{\sqrt{4\pi}}\int_b\ud\omega\left(a(\omega)e^{-i(k_c-k)z}+h.c.\right),\\
p(z)&=&-\frac{i}{\sqrt{4\pi}}\int_b\ud\omega\left(a(\omega)e^{-i(k_c-k)z}-h.c.\right)
\end{eqnarray}
\end{subequations}
with commutation relations $[x(z),p(z')]=ic\delta(z-z')$ where the
delta function has to be understood to have a width on the order
of $c/b$. Since we assumed that $\Omega\ll b$, the time it takes
for such a fraction of the pulse to cross the ensemble is much
smaller than the Larmor period $1/\Omega$. During the interaction
with one of these spatially localized modes the atomic state does
not change appreciable and we can simplify the interaction
operator to $V=\hbar\kappa (JT)^{-1/2}J_zp(0)$ where
$J_z=\sum_iF^{(i)}_z$ and we assumed that the ensemble is located
at $z=0$ and changed to a frame rotating at the carrier frequency
$\omega_c$.

A last approximation concerns the description of the atomic spin
state. Initially the sample is prepared in a coherent spin state
with maximal polarization along $x$, i.e. in the eigenstate of
$J_x$ with maximal eigenvalue $J$. We can thus make use of the
Holstein-Primakoff approximation \cite{K} which allows to describe
the spin state as a Gaussian state of a single harmonic
oscillator. The first step is to express collective step up/down
operators (along $x$), $J_{\pm}=J_y\pm iJ_z$, in terms of bosonic
creation and annihilation operators, $[b,b^\dagger]=\openone$, as
\begin{equation*}
J_+=\sqrt{2J}\sqrt{\openone-b^\dagger b/2J}\,b,\quad
J_-=\sqrt{2J}b^\dagger\sqrt{\openone-b^\dagger b/2J}.
\end{equation*}
It is easily checked that these operators satisfy the correct
commutation relations $[J_+,J_-]=2 J_x$ if one identifies
$J_x=J-b^\dagger b$. The fully polarized initial state thus
corresponds to the ground state of the harmonic oscillator. Note
that this mapping is exact. Under the condition that
\mbox{$\langle b^\dagger b\rangle\ll J$} one can approximate
\mbox{$J_+\simeq\sqrt{2J}b,\,J_-\simeq\sqrt{2J}b^\dagger$} and
therefore \mbox{$J_z\simeq-i\sqrt{J/2}(b-b^\dagger)$}. Introducing
atomic quadratures \mbox{$X=(b+b^\dagger)/\sqrt{2}$} and
\mbox{$P=-i(b-b^\dagger)/\sqrt{2}$} finally yields the desired
expression for the interaction \mbox{$V=\hbar\kappa T^{-1/2}
Pp(0)$}.

In terms of atomic quadratures the free Hamiltonian for atoms is
$H=\hbar\Omega/2(X^2+P^2)$. In the frame rotating at the carrier
frequency the action of $H_{li}$ on the light quadratures
$x(z),p(z)$ is simply
\mbox{$i/\hbar[H_{li},x(z)]=-c\parder{z}x(z)$} and likewise for
$p(z)$.

\paragraph*{Noise effects} The antihermitian part of the level
shift operator \eqref{levelshift} describes depumping of ground
state population and photon absorption. The effect of the latter
process can - as far as it concerns the performance of the
teleportation protocol - be treated on equal footing with mode
mismatch and finite detector efficiency. This is done in section
\ref{noise}. Loss of ground state population on the other hand
will eventually cause degrading of atomic polarization due to
spontaneous emission events. For a single atom the dominating term
describing this process stems from the scalar part of the
polarizability and is given by
$V_\mathrm{loss}=i\hbar\eta\openone/4T$ where
\mbox{$\eta=N_{ph}a_0\omega_c\Gamma
d^2/2\hbar\Delta^2\epsilon_0Ac=N_{ph}a_0\sigma\Gamma^2/4A\Delta^2$}.
It is possible to have $\eta\ll 1$ and at the same time a large
coupling $\kappa\simeq 1$. For a thermal cloud of atoms an
additional source of decoherence are light assisted collisions
which in fact dominate the decay process. Assuming a transverse
relaxation at an overall rate $\beta/T$ with $\beta\precsim .2$
the exponential decay during the interaction can to a good
approximation be treated linearly which leads to equations
\eqref{noisyatomstate}. The Langevin noise operators $f_{X,P}$ can
in principle be derived by a microscopic model as is done in
\cite{DCZP} for dephasing due spontaneous emission.


\section{Back action and input modes}\label{backaction}

\paragraph*{Back action} We evaluate here the input/output relations
(\ref{xcosout},\ref{xsinout}) for the cosine and sine components
of the in-phase field. For the former we take equation
(\ref{xlight}) at $\xi=c\tau,\,t=T$, multiply by
$\sqrt{2/T}\,\cos(\Omega\tau)$ and integrate over $\tau$ from $0$
to $T$. Using equations (\ref{Xsol}, \ref{Psol}) and the
approximate orthogonality of $\cos(\Omega\tau)$ and
$\sin(\Omega\tau)$ one finds
\begin{eqnarray*}
\x{c}{out}&=&\x{c}{in}+\frac{\kappa}{\sqrt{2}}\pa{in}+\\
&&+\frac{\sqrt{2}\kappa^2}{T^{3/2}}\int_0^T\ud\tau\int_0^\tau\ud\tau'[\cos(\Omega\tau)^2\sin(\Omega\tau')\bar{p}(c\tau',0)-\\
&&-\cos(\Omega\tau)\sin(\Omega\tau)\cos(\Omega\tau')\bar{p}(c\tau',0)].
\end{eqnarray*}
After interchanging the order of integration,
\mbox{$\int_0^T\ud\tau\int_0^\tau\ud\tau'\rightarrow\int_0^T\ud\tau'\int_{\tau'}^{T}\ud\tau$}
one can perform the integration over $\tau$. Neglecting all terms
of order $n_0^{-1}$ or less where $n_0=\Omega T\gg 1$ one finds
\begin{equation*}
\x{c}{out}\!=\!\x{c}{in}\!+\!\frac{\kappa}{\sqrt{2}}\pa{in}\!+\!\frac{\sqrt{2}\kappa^2}{T^{3/2}}\int_0^T\ud\tau\frac{T-\tau}{2}\sin(\Omega\tau)\bar{p}(c\tau,0).
\end{equation*}
The last term represents back action of light onto itself. It can
be expressed as a sum of two terms, one proportional to
$\p{s}{in}$ and another one proportional to
\begin{eqnarray}\label{back}
\p{s,1}{in}=\sqrt{3}\left(\frac{2}{T}\right)^{\!3/2}\!\!\!\int_0^T\ud\tau\!\left(\frac{T}{2}-\tau\right)\sin(\Omega\tau)\bar{p}(c\tau,0).&\quad&
\end{eqnarray}
It is easily verified that the back action mode defined by this
equation and the corresponding expression for $\x{s}{back}$ is
canonical
$[\x{s,1}{in},\p{s,1}{in}]=i[1-\mathcal{O}(n_0^{-2})]\simeq i$ and
independent from all the other modes introduced so far, f.e.
$[\x{s}{in},\p{s,1}{in}]=\mathcal{O}(n_0^{-2})\simeq 0$. The
variance is thus $(\Delta\p{s,1}{in})^2=1/2$. Repeating the
calculation for $\x{s}{out}$ with appropriate replacements and a
definition of $\p{c,1}{in}$ analogous to equation (\ref{back})
finally yields equations (\ref{xcosout},\ref{xsinout}).

In a similar way input-output relations for the back action modes
itself are derived. In particular for the in phase components one
finds
\begin{subequations}\label{xback}
\begin{eqnarray}
\x{c,1}{out}=\x{c,1}{in}
&-&\frac{1}{\sqrt{3}}\left(\frac{\kappa}{2}\right)^2\p{s}{in}+\frac{1}{\sqrt{15}}\left(\frac{\kappa}{2}\right)^2\p{s,2}{in},\quad\quad\\
\x{s,1}{out}=\x{s,1}{in}
&-&\frac{1}{\sqrt{3}}\left(\frac{\kappa}{2}\right)^2\p{c}{in}+\frac{1}{\sqrt{15}}\left(\frac{\kappa}{2}\right)^2\p{c,2}{in}.\quad\quad
\end{eqnarray}
\end{subequations}
In both equations the third terms on the right hand side describe
contributions of second order back action modes defined by
\begin{equation*}
\p{s,2}{in}=6\left(\frac{10}{T^5}\right)^{\!1/2}\!\!\!\int_0^T\ud\tau\!\left(\frac{T^2}{6}-T\tau+\tau^2\right)\sin(\Omega\tau)\bar{p}(c\tau,0)
\end{equation*}
and similarly for $\x{s,2}{in}$ and the cosine component. These
modes are again canonical and independent. As a sidemark we note
that, formally, it is possible to define scattering modes of
arbitrary order whose mode functions are given in general by
products of Legendre polynomials and $\cos(\Omega t)\,[\sin(\Omega
t)]$ resulting in a hierarchy of input-output relations similar to
\eqref{xback}.

\paragraph*{Input state}  The input field, propagating along the
positive $y$ direction and polarized along $z$ (see figure
\ref{scheme}), is described by operators
$[b(\omega),b^\dagger(\omega')]=\delta(\omega-\omega')$ in
frequency space and $[\hat{y}(y),\hat{q}(y')]=ic\delta(y-y')$ in
real space. ($\hat{y}$ is the quadrature operator for the field
in-phase and $y$ on the other hand is the position along the
$y$-direction.) In analogy to equation \eqref{localizedmodes}
these bases are connected via
\begin{eqnarray*}
\hat{y}(y)&=&\frac{1}{\sqrt{4\pi}}\int_b\ud\omega\left(b(\omega)e^{-i(k_c-k)y}+h.c.\right),\\
\hat{q}(y)&=&-\frac{i}{\sqrt{4\pi}}\int_b\ud\omega\left(b(\omega)e^{-i(k_c-k)y}-h.c.\right).
\end{eqnarray*}
As shown in section \ref{protocol} we can teleport the mode
\begin{equation*}
y=\frac{1}{\sqrt{2}}\left(\y{sin}+\q{cos}\right),\quad
q=-\frac{1}{\sqrt{2}}\left(\y{cos}-\q{sin}\right)
\end{equation*}
and the claim is that this corresponds to a pulse at the upper
sideband $\omega_c+\Omega$. Using the definitions of cosine and
sine modes \eqref{cosine} as well as
$\hat{\bar{y}}(c\tau,0)=\hat{y}(-c\tau,0)$ and the same for
$\hat{q}(y)$ we have
\begin{eqnarray*}
y&=&\frac{1}{\sqrt{T}}\int_0^T\ud\tau\left[\,\sin(\Omega\tau)\hat{\bar{y}}(c\tau,o)+\cos(\Omega\tau)\hat{\bar{q}}(c\tau,0)\right]\\
&=&\frac{-i}{\sqrt{4\pi T}}
\int_0^T\ud\tau\int_b\ud\omega\Bigl[b(\omega)e^{i(\omega_c+\Omega-\omega)\tau}-h.c.\Bigr]\\
q&=&-\frac{1}{\sqrt{T}}\int_0^T\ud\tau\left[\,\cos(\Omega\tau)\hat{\bar{y}}(c\tau,o)-\sin(\Omega\tau)\hat{\bar{q}}(c\tau,0)\right]\\
&=&\frac{-1}{\sqrt{4\pi T}}
\int_0^T\ud\tau\int_b\ud\omega\Bigl[b(\omega)e^{i(\omega_c+\Omega-\omega)\tau}+h.c.\Bigr].
\end{eqnarray*}
To explicitly see that this corresponds to a pulse centered at the
upper sideband it is convenient to change to a more precise model
by replacing the $1/\sqrt{T}$ factor, which is just the pulse's
slowly varying amplitude function in a simple square well
approximation, by a function $A(\tau)$ of dimension $s^{-1/2}$
normalized such that $\int_0^T\ud\tau|A(\tau)|^2=1$. Its Fourier
transform $A(\omega)=\frac{1}{\sqrt{2\pi}}\int_0^T\ud\tau
A(\tau)\exp(i\omega\tau)$ is centered at zero and has a width
$1/T=\Delta\omega\ll\Omega$ in accord with our condition
$1\ll\Omega T$. Replacing now $1/\sqrt{T}$ by $A(\tau)$ (of course
inside the integral over $\tau$) in the expressions for $y$ and
$q$ yields
\begin{eqnarray*}
y&=&\frac{-i}{\sqrt{4\pi}} \int_0^T\ud\tau A(\tau)
\int_b\ud\omega\Bigl[b(\omega)e^{i(\omega_c+\Omega-\omega)\tau}-h.c.\Bigr]\\
&=&\frac{-i}{\sqrt{2}}\int_b\ud\omega\Bigl[A(\omega_c+\Omega-\omega)b(\omega)-h.c.\Bigr],\\
q&=&\frac{-1}{\sqrt{2}}\int_b\ud\omega\Bigl[A(\omega_c+\Omega-\omega)b(\omega)+h.c.\Bigr].
\end{eqnarray*}
This is evidently a mode whose spectral mode function is the same
as the classical pulse but is centered at $\omega_c+\Omega$.


\section{Feedback}\label{feedback}

The feedback in continuous variable quantum teleportation is
sometimes described by equations equivalent to
\eqref{finalstateatoms1} but with a classical random variable
describing the measurement outcome in place of the operators
corresponding to the chosen displacement, which - though giving
the right result - is mathematically questionable. We point out
that relations \eqref{finalstateatoms1} hold \textit{stricto
sensu} as operator identities. This is true for mixed and even for
non-gaussian states, as we will show below.

Consider a bipartite system of $N+1$ modes and denote the first
mode as system $A$ and the remaining $N$ modes as system $B$. Let
the state of the compound system be given by $\rho_{AB}$. Our aim
here is to describe protocols which consist of the following
steps:
\paragraph*{Measurement} On system $B$ a set of commuting
observables $\{\hat{r}_1,\ldots,\hat{r}_N\}$ is measured where
each of the operators $\hat{r}_i$ is either $x_i$ or $p_i$, one of
the quadratures of mode $i$ in $B$. Let the corresponding
measurement outcomes $r_i$ be arranged in a vector
$\vec{R}^B=(r_1,\ldots,r_N)$. With the eigenvalue equation
$\hat{r}_i|r_i\rangle_B=r_i|r_i\rangle_B$, where $|r_i\rangle_B$
is the generalized eigenstate of $\hat{r}_i$, the normalized state
of system $A$ conditioned on the measurement outcomes is
$$
\rho_A^{(1)}(\vec{R})={}_B\langle
r_1,\ldots,r_N|\rho_{AB}|r_1,\ldots,r_N\rangle_B/p(\vec{R}).
$$
$p(\vec{R})$ is the probability to get the measurement outcomes
$\vec{R}$ and is normalized as $\int\ud^N\! r p(\vec{R})=1$.
\paragraph*{Feedback} Depending on the measurement outcomes
system $A$ is then displaced in $x_A$ and $p_A$ by an amount
$\vec{R}\vec{g}_x^T$ and $\vec{R}\vec{g}_p^T$ respectively where
$\vec{g}_{x(p)}$ are any real $N$ dimensional (row) vectors
determining the strength with which each outcome is fed back into
system $A$. In teleportation protocols these coefficients are
usually referred to as gains. The state of system $A$ after the
feedback operation is then
$$
\rho_A^{(2)}(\vec{R})=D_A^\dagger\rho_A^{(1)}(\vec{R})D_A.
$$
\mbox{$D_A=D_A(\vec{R}\vec{g}_x^T,\vec{R}\vec{g}_p^T)\doteq\exp(i\vec{R}\vec{g}_x^Tp_A)\exp(-i\vec{R}\vec{g}_p^Tx_A)$}
is the unitary displacement operator effecting the desired
transformations $D_Ax_AD_A^\dagger=x_A+\vec{R}\vec{g}_x^T$ and
\mbox{$D_Ap_AD_A^\dagger=p_A+\vec{R}\vec{g}_p^T$}.
\paragraph*{Ensemble average} On average over all
measurement outcomes, weighted with their respective
probabilities, the state of system $A$ is
$$
\bar{\rho}_A=\int\ud^N\!rp(\vec{R})\rho_A^{(2)}(\vec{R}).
$$
Combining this with the expressions for $\rho_A^{(2)}$ and
$\rho_A^{(1)}$ above one can express
\begin{eqnarray}
\bar{\rho}_A&=&\int\ud^N\!rD_A^\dagger{}_B\langle
r_1,\ldots,r_N|\rho_{AB}|r_1,\ldots,r_N\rangle_BD_A\nonumber\\
&=&\int\ud^N\!r{}_B\langle
r_1,\ldots,r_N|D_{AB}^\dagger\rho_{AB}D_{AB}|r_1,\ldots,r_N\rangle_B\nonumber\\
&=&\mathrm{tr}_B\{D_{AB}^\dagger\rho_{AB}D_{AB}\}\label{averagestate}
\end{eqnarray}
where the trace in the last line is now taken with respect to both
system $A$ and $B$. In going from the first line to the second
line we made use of the identity
\begin{equation}\label{id}
|r_1,\ldots,r_N\rangle_BD_A=D_{AB}|r_1,\ldots,r_N\rangle_B
\end{equation}
with the unitary operator $D_{AB}$ defined as $
D_{AB}=D_{AB}(\hat{\vec{R}}\vec{g}_x^T,\hat{\vec{R}}\vec{g}_p^T)\doteq\exp(i\hat{\vec{R}}\vec{g}_x^Tp_A)\exp(-i\hat{\vec{R}}\vec{g}_p^Tp_A)
$ where $\hat{\vec{R}}=(\hat{r}_1,\ldots,\hat{r}_N)$ is now the
vector of \textit{operators} $\hat{r}_i^B$ and $D_{AB}$ acts on
\textit{both} systems $A$ and $B$. Note that identity \eqref{id}
is valid only for commuting observables $\hat{r}_i^B$. The
resulting equation \eqref{averagestate} is the key  point in this
consideration.
\paragraph*{Observables in the ensemble average} Consider
finally the mean of, for example, $x_A$ after the measurement and
feedback procedure, i.e. with respect to the ensemble averaged
state $\bar{\rho}_A$. It is given by
\begin{eqnarray*}
\langle x_A\rangle
&=&\mathrm{tr}_A\{x_A\bar{\rho}_A\}=\mathrm{tr}_{AB}\{D_{AB}x_AD_{AB}^\dagger\rho_{AB}\}\\
&=&\mathrm{tr}_{AB}\{(x_A+\hat{\vec{R}}\vec{g}_x^T)\rho_{AB}\}.
\end{eqnarray*}
From this identity and the corresponding expression for $\langle
p_A\rangle$, which both are true for all initial states
$\rho_{AB}$, we can deduce the operator identities
$$
x_A^{\mathrm{fin}}=x_A+\hat{\vec{R}}\vec{g}_x^T,\quad
p_A^{\mathrm{fin}}=p_A+\hat{\vec{R}}\vec{g}_p^T
$$
where $x_A^{\mathrm{fin}},p_A^{\mathrm{fin}}$ describe the final
state of system $A$ in the Heisenberg picture and means have to be
taken with respect to the unchanged initial state of both systems
$A$ and $B$. If $\rho_{AB}$ is a pure Gaussian state the last two
equations fully determine the final state $\bar{\rho}_A$. This was
used in equation \eqref{finalstateatoms1}. Note that these
considerations are easily extended to situations in which system
$A$ consists of more than one mode.



\begin{thebibliography}{54}

\bibitem{PHOTONS} D. Bouwmeester et al., Nature \textbf{390}, 575
(1997), D. Fattal et al., Phys. Rev. Lett. \textbf{92}, 037904
(2004), I. Marcikic et al., Nature \textbf{421}, 509 (2003), Y.-H.
Kim et al., Phys. Rev. Lett. \textbf{86}, 1370 (2001), D. Boschi
et al., Phys. Rev. Lett. \textbf{80}, 1121 (1998), R. Ursinn et
al., Nature \textbf{430}, 849 (2004), J.-W. Pan et al., Phys. Rev.
Lett. \textbf{86}, 4435 (2001), A. Furusawa et al., Science
\textbf{282}, 706 (1998), T.C. Zhang et al., Phys. Rev. A
\textbf{67}, 033802 (2003), W.P. Bowen et al., Phys. Rev. A
\textbf{67}, 032302 (2003), H. Yonezawa et al., Nature
\textbf{431}, 430 (2004), N. Takei et al., Phys. Rev. Lett.
\textbf{94} 220502 (2005)

\bibitem{ATOMS} M.Riebe et al., Nature \textbf{429}, 734 (2004),
M.D. Barrett et al., Nature \textbf{429}, 737 (2004)

\bibitem{JSCFP} B. Julsgaard, J. Sherson, J.I. Cirac, J. Fiurasek, E.S.
Polzik, Nature \textbf{432}, 482 (2004)

\bibitem{BP} S. Braunstein and A. Pati, \textit{Quantum Information with Continuous
Variables}, Kluwer, New York, 2003

\bibitem{HCWL} M.J. Holland, M.J. Collett, D.F. Walls, M.D.
Levenson, Phys. Rev. A \textbf{42}, 2995 (1990)

\bibitem{PRG} J.P. Poizat, J.F. Roch, P. Grangier, Ann. Phys. Fr. \textbf{19}, 265 (1994)

\bibitem{JKP} B. Julsgaard, A. Kozhekin, E. S. Polzik, Nature
\textbf{413}, 400 (2001)

\bibitem{V} L. Vaidman, Phys. Rev. A, \textbf{49}, 1473 (1994)

\bibitem{BK} S.L. Braunstein, H.J. Kimble, Phys. Rev. Lett. \textbf{80}, 869
(1998)

\bibitem{KMB} A. Kuzmich, L. Mandel, N.P. Bigelow, Phys. Rev. Lett.
\textbf{85}, 1594 (2000)

\bibitem{GSM} J.M. Geremia, J.K. Stockton, H. Mabuchi, Science \textbf{304}, 270
(2004)

\bibitem{OF} T. Opatrny, J. Fiurasek, Phys. Rev. Lett. \textbf{95}, 053602 (2005)

\bibitem{MF} L. Mista, R. Filip,  Phys. Rev. A \textbf{71} 032342
(2005)

\bibitem{KMPP} A. Kuzmich, K. M\o lmer, E.S. Polzik, Phys. Rev. Lett. \textbf{79},
4782 (1997), E.S. Polzik, Phys. Rev. A \textbf{59}, 4202 (1999)

\bibitem{KBM} A. Kuzmich, N.P. Bigelow, L. Mandel, Europhys. Lett. \textbf{42}, 481
(1998)

\bibitem{TMW} L.K. Thomsen, S. Mancini, H.M. Wiseman, Phys. Rev. A \textbf{65},
061801 (2002)

\bibitem{DCZP} L.-M. Duan, J.I. Cirac, P. Zoller, E.S. Polzik,
Phys. Rev. Lett. \textbf{85}, 5643 (2000)

\bibitem{KMP} A. E. Kozhekin, K. M\o lmer, E.S. Polzik, Phys. Rev. A \textbf{62}, 033809 (2000)


\bibitem{JSSP} B. Julsgaard, C. Schori, J.L. S\o rensen, E.S. Polzik., Quant. Inform. and Comp.,
spec. issue \textbf{3}, 518 (2003)

\bibitem{J} B. Julsgaard, \textit{Entanglement and Quantum
Interactions with Macroscopic Gas Samples}, PhD Thesis, October
2003, Aarhus University, Denmark,
http://www.nbi.dk/$\sim$julsgard/

\bibitem{SJP} J. Sherson, B. Julsgaard, E.S. Polzik,
quant-ph/0408146

\bibitem{KMSJP} D.V. Kupriyanov, O.S. Mishina, I.M. Sokolov, B. Julsgaard,
E.S. Polzik, quant-ph/0411083

\bibitem{K} C. Kittel, \textit{Quantum Theory of Solids}, Wiley, New York, 1987

\bibitem{SD} A. Silberfarb, I.H. Deutsch, Phys. Rev. A \textbf{68}, 13817 (2003)

\bibitem{MM} L.B. Madsen, K. M\o lmer, Phys. Rev. A \textbf{70}, 052324 (2004)

\bibitem{CS} C.M. Caves, B.L. Schumaker, Phys. Rev. A \textbf{31},
3068 (1985), B.L. Schumaker, C.M. Caves, Phys. Rev. A \textbf{31},
3093 (1985)

\bibitem{GKLC} G. Giedke, B. Kraus, M. Lewenstein, J.I. Cirac, Phys. Rev. A \textbf{64}, 052303 (2001)

\bibitem{BFK} S.L. Braunstein, H.J. Kimble, C.A. Fuchs, J. Mod. Opt. \textbf{47}, 267 (2000)

\bibitem{HWPC} K. Hammerer, M.M. Wolf, E.S. Polzik, J.I. Cirac, Phys. Rev. Lett. \textbf{94}, 150503 (2005)

\bibitem{GWKWC} G. Giedke, M. M. Wolf, O. Krüger, R. F. Werner, J. I. Cirac
Phys. Rev. Lett. \textbf{91}, 107901 (2003)

\bibitem{GLP} P. Grangier, J.A. Levenson, J.P. Poizat, Nature
\textbf{396}, 537 (1998)

\bibitem{GC} G. Giedke, J.I. Cirac, Phys. Rev. A \textbf{66},
032316 (2002)

\bibitem{H} W. Happer, Rev. Mod. Phys. \textbf{44}, 169 (1972)



\end{thebibliography}
\end{document}